\documentclass[12pt,reqno]{article}

\usepackage{amssymb}
\usepackage{amsmath}
\usepackage{amsthm}
\usepackage{amsfonts}
\usepackage{enumerate}
\usepackage{url}
\usepackage{graphicx}
\usepackage{textcomp}
\usepackage{bm}
\usepackage{apacite}
\usepackage{setspace}

\newcommand{\vv}{\overrightarrow}

\begin{document}

\title{Elongation and stability of a linear dune}

\author{O. Rozier$^1$, C. Narteau$^1$, C. Gadal$^1$, P. Claudin$^2$ \\ and S. Courrech du Pont$^3$}

\date{}

\maketitle

\vskip 1cm

\begin{small}
{\setstretch{0.8}
\textbf{Key Points:}
\begin{itemize}
\item Linear dunes with a finger-like shape are identified as an elementary dune type.
\item The longitudinal mass balance governs the elongation and stability of a linear dune.
\item Dune morphodynamics is controlled by reversing winds and by the minimum dune size at the tip.
\end{itemize}
}
\end{small}

\vskip 1cm

\begin{abstract}
Compared to barchan dunes, the morphodynamics of linear dunes that elongate on a non-erodible bed is barely investigated by means of laboratory experiments or numerical simulations. Using a cellular automaton model, we study the elongation of a solitary linear dune from a sand source and show that it can reach a steady state. This steady state is analyzed to understand the physical processes at work along the dune. Crest reversals together with avalanche processes control the shape of transverse sections. Dune width and height decrease almost linearly with distance downstream until the minimum size for dunes. This is associated with a constant sand loss along the dune, which eventually compensates for the sediment influx and sets the dune length. This sand budget is discussed to distinguish an elongating linear dune from a barchan dune and to explain the complexity of linear dune fields in nature.
\end{abstract}

\vskip 1cm

\noindent\rule{\textwidth}{0.5pt}

{\setstretch{0.8}

\noindent
{\footnotesize $^1$ Institut de Physique du Globe de Paris, UMR 7154 CNRS, Universit\'e de Paris, Paris, France.}

\vskip 0.2cm

\noindent
{\footnotesize $^2$ Physique et M\'ecanique des Milieux H\'et\'erog\`enes, UMR 7636 CNRS, ESPCI Paris, PSL Research University -- Sorbonne Universit\'e --  Universit\'e de Paris, Paris, France.}

\vskip 0.2cm

\noindent
{\footnotesize $^3$ Laboratoire Mati\`ere et Syst\`emes Complexes, UMR 7057 CNRS, Universit\'e de Paris, Paris, France.}

}

\newpage

\section*{Plain Language Summary}
Given the seasonal cycle, multidirectional wind regimes are highly prevalent in modern sand seas where linear dunes are the most frequent dune type. They vary in shape and size but they are all characterized by linear ridges extending over long distances. Here we study the morphodynamics of linear dunes that elongate from a sand source under the action of reversing winds. We show that the sand loss increases with the length of the elongating linear dune, which eventually converges to a steady state. We relate the dimensions and the shape of the dune at equilibrium to the sediment influx and the wind properties. The characterization of this elementary dune type is an essential step towards a better understanding of the interplay between winds and complex dune fields in nature. This is important for Earth in a context of climate change, but also for Mars and Titan, Saturn's largest moon, where direct wind measurements are not available to date.

\noindent\rule{\textwidth}{0.5pt}

\section{Introduction}

Elongating linear dunes are individual sand ridges aligned near the resultant transport direction (Figure~\ref{fig: 1}a). Also referred to as seif dunes \cite{Lanc82,Tsoa82} or silks \cite{Main78} when they are sinuous, these finger-like structures are widespread on Earth and other planetary bodies. They develop on non-erodible beds submitted to multidirectional flow regimes thanks to the deposition at the dune tips of the sediment transported along the crests \cite{Cour14,Gao15a}. Sometimes the upstream source of sediment is fixed, like for lee dunes elongating behind a topographic obstacle \cite{Tsoa89}. In such a situation, there is no lateral migration of  the dune body, which can preserve its shape over tens of kilometers \cite{Lu17,Luca14}. Understanding the sediment budget along these longitudinal dunes and the conditions leading to morphodynamic stability is key to assess time and length scales associated with the mechanism of dune growth by elongation.

An intrinsic feature of elongating dunes is that they grow and form under the combined effect of winds blowing successively on either side of the crest. This alternation of reversing winds causes deposition along the flanks and at the tip of the dunes. The subsequent sedimentary structures have been documented in cold and hot aeolian dunes \cite{Bris00,Bris10} but have never been quantitatively related to the surface processes and the overall dune morphodynamics. 

Assessing the variability of linear dune fields first requires an understanding of how an individual linear dune forms and evolves \cite{Livi89}. Furthermore, recent advances show that a complex dune  morphology is often the result of interacting elementary bedforms. For instance, one can interpret star dunes as a combination of individual elongating dunes \cite{Zhan12}. Similarly, raked linear dunes can be described as linear dunes with superimposed regularly spaced barchan dunes migrating in an oblique direction \cite{Lu17}. This reductive approach implies the accurate identification of elementary dune features. This is the main objective of this numerical study in regards to elongating linear dunes. 

The next section provides an outline of the numerical model and describes the different setups as well as the associated parameter space. In section~\ref{sec: morpho}, we show that an elongating linear dune can reach a steady state. Subsequently, we determine the mechanisms that govern this dynamic equilibrium under various conditions. Finally, we discuss the properties and stability of elongating dunes with respect to their counterparts in a unidirectional wind regime, namely the barchan dunes.   

\section{Methods}
\label{sub:setup}
Numerical simulations are performed using a cellular automaton dune model that accounts for feedback mechanisms between the flow and the bed topography \cite{Nart09,Rozi14}. The saturated sediment flux  depends on a threshold shear stress $\tau_1$ for motion inception. All the numerical results are expressed in units of $\{l_0,\,t_0\}$, which are the characteristic length and time scales of the model \cite{Nart09}. They relate to the most unstable wavelength $\lambda_{\rm max}$ for the formation of dunes and to the saturated sand flux $q_{\rm sat}$. When $\tau_1$ tends to zero, $\lambda_{\rm max} \sim 40 \; l_0$ and $q_{\rm sat} \sim 0.23 \; l_0^2/t_0$.

In all simulations, we set an asymmetric bidirectional wind regime of period $T$. Over a wind cycle, two winds of the same strength blow alternatively with a divergence angle $\theta=120^{\rm o}$. The duration of the primary wind is twice that of the secondary wind, resulting in a mass transport ratio $N=2$ on a flat bed. The two winds are oriented such that the dune elongates along the main axis of the cellular space of the model. The angle $\alpha$ between the elongation direction and the primary wind is $41 \pm 1^{\rm o}$, in agreement with the predictions in \citeA{Cour14} and \citeA{Gao15a}:  $\tan \alpha = \sin \theta / ( \sqrt{N} + \cos \theta ) $.

We consider two different setups to investigate the physical mechanisms governing the elongation and the stability of linear dunes. In the first setup, hereafter the ``injection setup", the simulated field is a corridor with open boundary conditions. Sediment is injected locally from a fixed circular source near the upstream end of the field at a constant volume rate $J_{\rm in}$ (see inset in Figure~\ref{fig: 1}b). Using the injection setup, the model becomes too much time and space consuming when simulating hugely elongated dunes. To analyze transverse sections over a wider range of dune sizes and wind conditions, we also simulate longitudinal sand piles with periodic boundary conditions. We refer to this setup as the ``infinite setup". Starting from a uniform square cross section at $t=0$, the sand pile reaches the characteristic shape of a reversing dune in a few wind cycles. After an elapsed time of ${\sim}10^4\;t_0$, we examine the shape  properties.

Keeping the same bidirectional wind regime in terms of parameters $\{\theta,\, N\}$, we vary the $\{\tau_1,\, T,\, J_{\rm in}\}$-values to explore how the transport threshold, the duration of the wind cycle and the sediment injection rate impact the morphodynamics of elongating linear dunes. We perform all shape measurements at the end of the secondary wind. For each simulation, we regularly estimate the volume $V$ and the length $L$ of the dune as well as the height $H$, width $W$, and area $S$ of all transverse sections. We compute the associated aspect ratio $\rho = H/W$ and shape ratio $\phi = S/(WH)$, which typically ranges between $1/2$ and $2/3$, two values that correspond to triangular and parabolic sections, respectively.  

We also measure the crest reversal distance $\Delta_{\rm c}$, i.e., the travel distance of the crest line between two alternate winds, perpendicularly to the elongation direction. For each transverse section, we focus on  the outflux $q_{\rm out}$ that escapes from the dune, and on the mean longitudinal sand flux ${\langle q \rangle}$, which is defined as the cumulative flux in the direction of elongation divided by the width of the section. All sand fluxes are computed by counting transitions of transport occurring within the entire cellular space \cite{Zhan14}.

\section{Stabilization and shape of an elongating linear dune}
\label{sec: morpho}
\begin{figure}
\begin{center}
\includegraphics[width=1.0\linewidth]{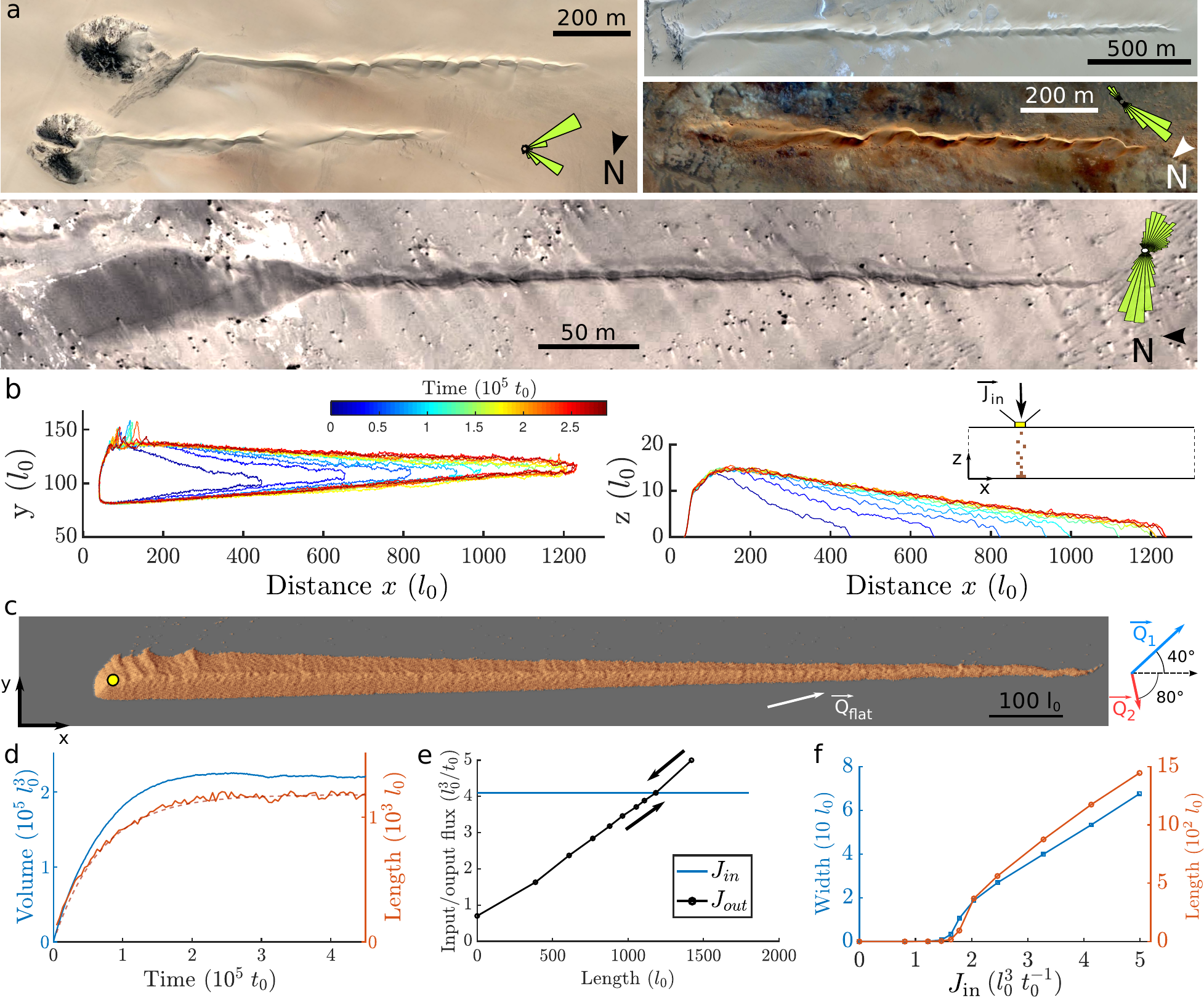}
\end{center}
\caption{
{\bf Formation, stabilization, and shape of an elongating linear dune.}
{\bf a:}
Elongating linear dunes and sand flux roses in terrestrial dune systems: 
(top) Niger
(16\textdegree52\textasciiacute{}N, 13\textdegree20\textasciiacute{}E 
and
18\textdegree21\textasciiacute{}N, 13\textdegree07\textasciiacute{}E),
(middle) Mauritania
(23\textdegree12\textasciiacute{}N, 10\textdegree50\textasciiacute{}W),
(bottom) China
(39\textdegree54\textasciiacute{}N, 94\textdegree09\textasciiacute{}E).
Imagery from Google Earth and Maxar.
{\bf b:} 
Vertical and horizontal contours of a simulated linear dune elongating from
a fixed sand source with a constant influx (see inset).
{\bf c:}
Steady-state dune shape.
Transport vectors $\protect\vv{Q}_1$ and  $\protect\vv{Q}_2$ of
primary and secondary winds and resultant $\protect\vv{Q}_{\rm flat}=\protect\vv{Q}_1+\protect\vv{Q}_2$ are shown in blue, red, and white, respectively. 
The black arrow is the predicted dune orientation. The sand source is depicted by the yellow disk.
{\bf d:} 
Volume and length of a linear dune as a function of time. The red dashed line shows the prediction.
{\bf e:} 
Stability diagram. The steady length is selected when the input ($J_{\rm in}$) and output ($J_{\rm out}$) fluxes balance.
{\bf f:} 
Length and maximum width of a steady linear dune with respect to sediment influx.
}
\label{fig: 1}
\end{figure}

Using the injection setup, dunes elongate only above a critical sediment influx that depends on the transport threshold and the wind regime. Below the critical influx, elongation is impeded by sediment loss in the injection area and frequent breakups of the dune body. Above, the linear dune elongates and eventually reaches a steady state (Figure~\ref{fig: 1}b-e). Figure~\ref{fig: 1}f shows the length and maximum width of the dune at steady state as a function of the sediment influx $J_{\rm in}$ when setting $T=300\;t_0$ and $\tau_1=0$. For this wind regime, dunes elongate when $J_{\rm in} \gtrsim 2 \; l_0^3\,t_0^{\rm -1}$. Above the onset of elongation, both the maximum length and width are linearly related to $J_{\rm in}$ in the range of investigated values (up to $J_{\rm in} = 5 \; l_0^3\,t_0^{\rm -1}$). In the next section, wind conditions remain identical and $J_{\rm in}$ is set to $4.1\; l_0^3\, t_0^{-1}$.


\subsection{Formation and length stabilization}
\label{sub:flux}
As soon as sediment begins to accumulate in the injection area, a sand pile forms and elongates under the action of successive winds (Figures~\ref{fig: 1}b and \ref{fig: 1}d). As in laboratory experiments \cite{Cour14}, the elongating linear dune has a finger-like structure at all times with a straight crest line, sharp boundaries, and a reversing slip face. The height and width of the cross sections in the injection area rapidly reach stationary values. From the source, the height and width of cross sections decrease almost linearly with distance up to the dune tip (see the triangular contours in Figure~\ref{fig: 1}b). As a result, the mean cross-sectional area of the elongating dune $\langle S \rangle = V / L$ is constant over time. 

\begin{figure}
\centering
\includegraphics[width=\textwidth,height=\textheight,keepaspectratio]{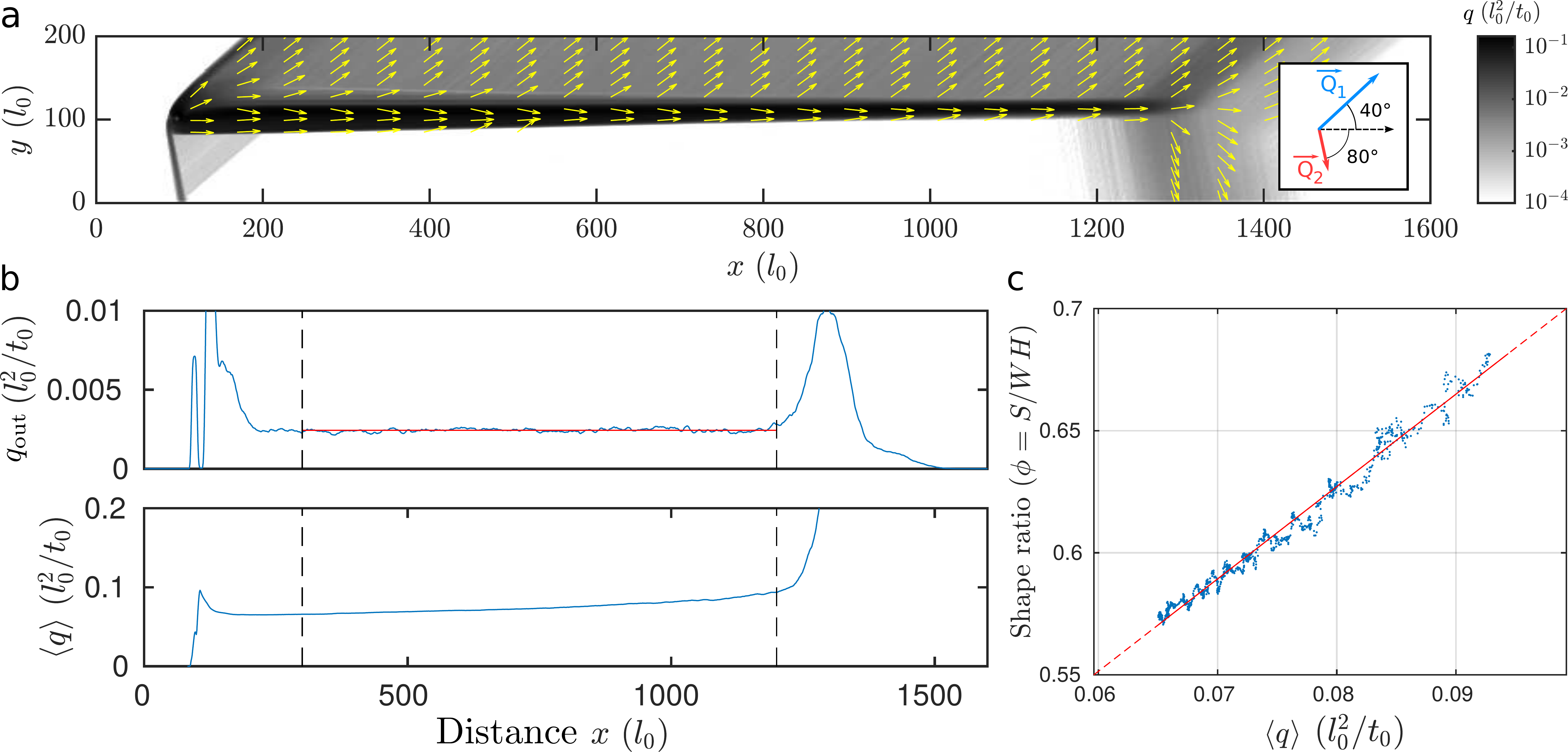}
\caption{
{\bf Sand fluxes along a steady linear dune.}
\textbf{a:}
Intensity (grayscale) and direction (yellow arrows) of the sand flux averaged over 10 wind cycles.
\textbf{b:}
Sediment loss $q_{\rm out}$ (top) and mean longitudinal sand flux $< q >$ (bottom) along the dune.
\textbf{c:}
Dune shape ratio $\Phi$ with respect to the mean longitudinal sand flux $< q >$.
The solid line shows the best linear fit.}
\label{fig: 2}
\end{figure}

The sand source holds in place the upstream end of the dune, which prevents the dune to migrate laterally. The direction of elongation is constant over time and parallel to the resultant transport direction at the crest (yellow arrows in Figure \ref{fig: 2}a). In our asymmetric bidirectional wind regime, this transport direction at the crest is oblique to the resultant drift direction on a flat sand bed (RDD, see transport vectors in Figure~\ref{fig: 1}c). This is due to a difference in speed-up between the two winds, according to the dune aspect ratio experienced by the wind. The elongation rate is maximum at the beginning of the simulation and decreases as the dune converges to its steady state (Figures~\ref{fig: 1}b and \ref{fig: 1}d). Crest reversals continuously modify the overall dune shape, but both dune length and volume reach stationary values when averaged over a few wind cycles (Figure \ref{fig: 1}d).

The elongating dune undergoes high sediment losses in the injection area and at the tip, but a  moderate outflux also escapes along the dune body from the lee side of the primary wind (Figures~\ref{fig: 2}a-b). This is because the dune alignment makes an angle with the RDD. Relatively to the crest, the RDD points towards the same side as the primary wind. Moreover, the direction of the primary wind is oblique to the dune crest, so that it experiences a smaller apparent aspect ratio than a perpendicular wind. In the absence of an apparent slip face, the lee side of the dune becomes a less efficient sand trap. This is observed in nature where a significant outflux can be emitted from the lee side of oblique winds (see wind streaks in Figure~\ref{fig: 1}a).

Therefore, under open boundary conditions the total sediment loss $J_{\rm out}$ increases with an increasing dune length, until it balances the influx coming from the injection area (Figures \ref{fig: 1}e-f). Interestingly, the intensity of the outflux along the dune body is uniform and stationary when averaged over a wind cycle (Figure~\ref{fig: 2}b). This property simplifies
the derivation of dune elongation, which is governed by the overall sediment budget. The conservation of mass for the entire dune of volume $V$ and length $L$ reads

\begin{equation}\label{eq:Vdyn}
\frac{\partial V}{\partial t} = J_{\rm in} - J_{\rm ex} - L\,q_{\rm out},
\end{equation}

\noindent
where $q_{\rm out}$ is the outflux along the dune and $J_{\rm ex}$ a constant volume rate that accounts for the extra sediment loss occurring at both extremities. Substituting $V$ by $ L \langle S \rangle$ and integrating equation \eqref{eq:Vdyn} yields

\begin{equation} \label{eq:L_t}
L = \displaystyle L_s \left( 1 - e^{-\frac{t}{\tau}} \right),
\end{equation}

\noindent
where $\tau = \langle S \rangle / q_{\rm out}$ is the characteristic time of stabilization and $L_s$ the steady length that reads

\begin{equation} \label{eq:L_S}
L_s = \left(  J_{\rm in} - J_{\rm ex} \right) / q_{\rm out}.
\end{equation}

\noindent
The relaxation equation~\eqref{eq:L_t} agrees with the numerical outputs (Figure~\ref{fig: 1}b). The observed values of $L_s \sim 1175\;l_0$ and $q_{\rm out} \sim 2.5 \times 10^{-2} \,l_0^2 \, t_0^{-1}$ give $J_{\rm ex} \sim 0.3 \; J_{\rm in}$.

The integrated flux $W {\langle q \rangle}$ over a transverse section verifies the equation of mass conservation at steady state:

\begin{equation} \label{ed:Qs_x}
  \frac{\partial \left(  W {\langle q \rangle}\right)}  {\partial x}  = -q_{\rm out}.
\end{equation}

\noindent
The mean longitudinal sand flux ${\langle q \rangle}$ is expected to depend on the shape of the cross section. As this shape barely changes along the dune, we assume a constant ${\langle q \rangle}$ independent of $x$ as a first approximation. This gives

\begin{equation}
 \frac{\partial W}{\partial x} = - \frac{q_{\rm out}}{\langle q \rangle}.
\end{equation}

\noindent
Hence, the constant outflux $q_{\rm out}$ explains the triangular contours of the dune (Figure~\ref{fig: 1}c). The linear decrease of width and height with the distance from the source implies ${\langle S \rangle} \propto {\langle W \rangle}^2 $. Combined with the linear dependency between the maximum width of the dune and the sediment influx (Figure~\ref{fig: 1}f), the characteristic time $\tau$ for the dune stabilization roughly goes like $J_{\rm in}^2$. 

In the limit of long dunes, for which $L q_{\rm out} \gg J_{\rm ex}$, the dune maximum length simply reads $L_s \simeq J_{\rm in} / q_{\rm out}$. We expect a constant sand loss $q_{\rm out}$ per unit length all along the dune at least as long as the dune remains simple (i.e., with a regular shape and free of superimposed bedforms). This property could be used to infer the sediment budget in the field.

We observe in Figure \ref{fig: 2}b that ${\langle q \rangle}$ is not strictly constant but slightly increases with the distance from the source. This is a consequence of the evolution of the dune shape, as shown by the linear relationship between the mean longitudinal sand flux ${\langle q \rangle}$ and the shape ratio $\phi$ (Figure \ref{fig: 2}c). Hereafter, we also use the infinite setup to precisely relate the dune morphodynamics and the associated sand fluxes.


\subsection{Steady-state morphology}
\label{sub:morpho}

\begin{figure}
\includegraphics[width=1.0\linewidth]{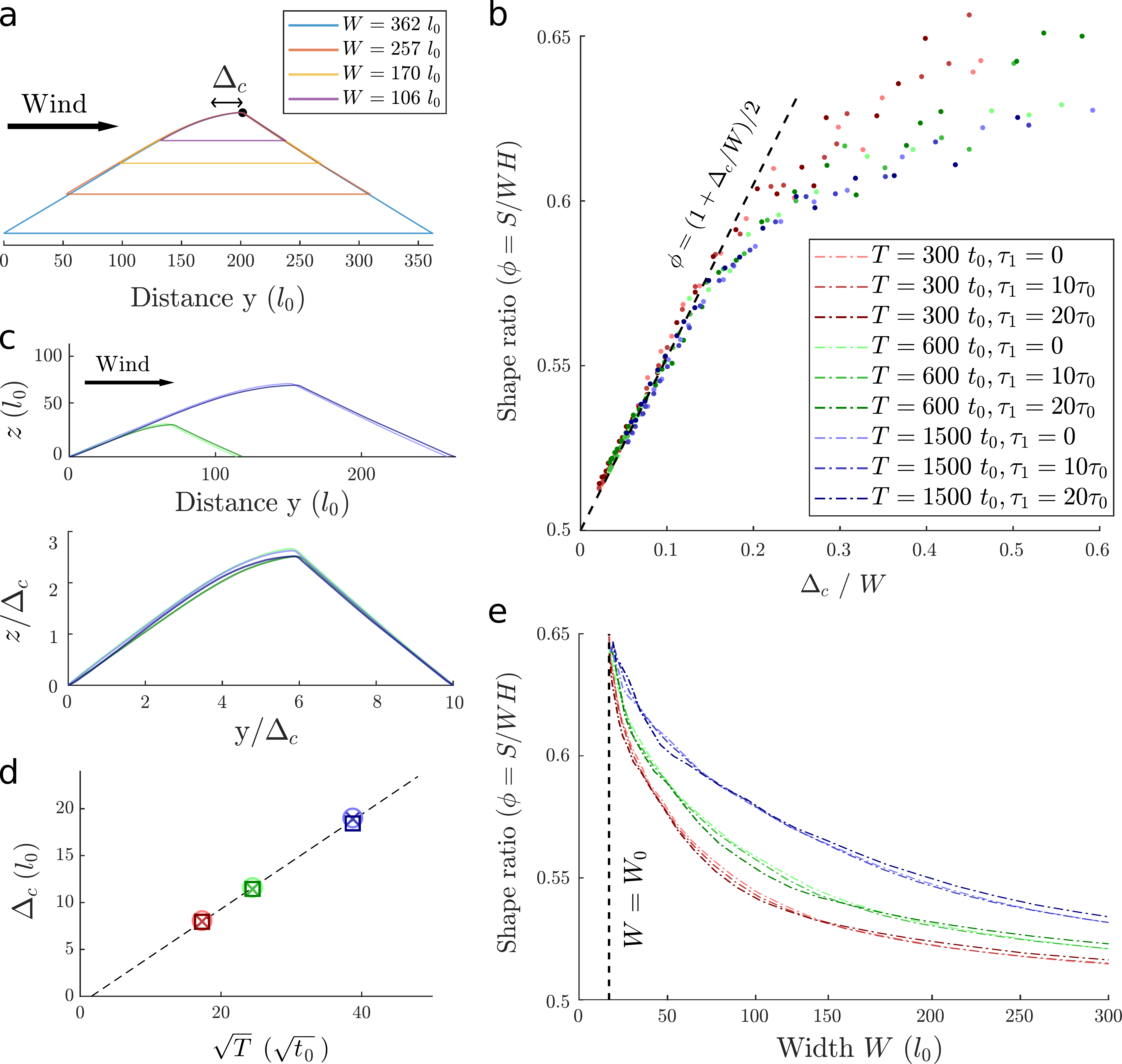}
\caption{
{\bf Shape of transverse sections along a steady linear dune.}
\textbf{a:} Transverse sections of different width shaped by the same wind regime of period $3000\,t_0$. Cross sections superimpose when matching the crest positions (black point).
\textbf{b:} Shape ratio $\phi$ with respect to the ratio $\Delta_{\rm c} / W$ for various $\{T, \tau_1 \}$.
\textbf{c:} Transverse sections of different widths having same ratio $W/\Delta_{\rm c} = 10$ (top). The elevation profiles overlap almost perfectly once normalized (bottom).
\textbf{d:} Crest reversal distance $\Delta_{\rm c}$ as a function of the square root of the wind period $\sqrt{T}$. Different symbols are for different values of $\tau_1$.
\textbf{e:} Shape ratio $\phi$ with respect to the width $W$.
From \textbf{b} to \textbf{e}, the same color code is used (inset in \textbf{b}), except in \textbf{c} where blue elevation profiles are for $T=3000\;t_0$.
}
\label{fig: 3}
\end{figure}

We first compare transverse sections of various widths under the same periodic wind regime as in Figures~\ref{fig: 1} and \ref{fig: 2}. Despite the asymmetry of the wind regime, sections have a rather symmetric shape with slip faces in the lower part of both sides. These lower parts are barely reworked by winds contrary to the upper area where the crest line moves back and forth. At the end of each period of constant wind orientation, the elevation profile of this reworked area resembles the central slice of a barchan dune. The crest reversal distance $\Delta_{\rm c}$ is constant for all dune sections, so that the elevation profiles of cross sections with different widths overlap when matching the crest positions (Figure~\ref{fig: 3}a). When the lower part is large compared to the reworked upper part, sections are well approximated by trapezoid of bases $W$ and $\Delta_{\rm c}$ and height $H$. The corresponding shape ratio can be expressed as

\begin{equation}
\label{eq:phi_scale}
\phi \sim \frac12 \left( 1 + \frac{\Delta_{\rm c}}{W} \right).
\end{equation}

Figure~\ref{fig: 3}b shows the shape ratio $\phi$ as a function of $\Delta_{\rm c} / W$ for three transport thresholds $\tau_1/\tau_0 =\{ 0, \, 10, \, 20 \} $ and three wind cycle durations $T/t_0= \{300, \,600,\, 1500 \}$. The largest dune sections satisfy the trapezoid shape equation~\ref{eq:phi_scale} when  $\Delta_{\rm c}/W<0.15$. This common behavior suggests that large cross sections develop a self-similar shape when widths and heights are normalized by $\Delta_{\rm c}$. Figure~\ref{fig: 3}c shows transverse sections with the same ratio $\Delta_{\rm c}/W = 0.1$ before and after normalization. Normalized profiles collapse onto a common profile in the lower part. At the crest, we find good agreement for different $T$-values but not for different $\tau_1$-values. Indeed, the aspect ratio of the reworked area decreases when the transport threshold increases. This is consistent with previous numerical results showing that barchan and transverse dunes flatten when the transport threshold increases \cite{Gao15b,Zhan10}. The same conclusion can be drawn from a linear stability analysis. In the limit of large dunes, the dune velocity is independent of the transport threshold but the growth rate decreases when the transport threshold increases (equations~2.12, ~2.13 in \citeA{Gada19}). Thus, on the one hand, the value of $\tau_1$ sets the aspect ratio of the upper part of the dune. On the other hand, when $\tau_1$ is fixed and $T$ varies, the cross-sectional area reworked during a time period $T$ is proportional to $\Delta_{\rm c} ^2$ and to $q_{\rm sat} T$. Accordingly, we find that $\Delta_{\rm c}$ is proportional to $\sqrt{T}$, whereas it is independent of $\tau_1$ (Figure~\ref{fig: 3}d).

The fact that $\Delta_{\rm c}$ is constant along the dune (see Figure~\ref{fig: 3}a) explains the increase of the mean sand flux $\langle q \rangle$ with an increasing shape ratio $\phi$ (Figure \ref{fig: 2}c). For a given cross section, the sand flux is found maximum in the mobile crest area of width ${\sim}\Delta_{\rm c}$. Thus, when the ratio $\Delta_{\rm c} /W$ and the shape ratio $\phi$ increase, the mean sand flux $\langle q \rangle$ increases accordingly.


While the shape of large cross sections is controlled by the reversing distance $\Delta_{\rm c}$, it is not the case for the small cross sections close to the dune tip (Figure \ref{fig: 3}b). Whatever the wind cycle duration $T$ and the transport threshold $\tau_1$, the shape ratio $\phi$ converges to a maximum value ${ \sim} 0.65$ for a minimum section width $W_0 \sim 15\;l_0$ (Figure \ref{fig: 3}e). This minimum length scale relates to the lower cut-off wavelength $\lambda_{\rm c}$ of the flat bed instability and to the minimum size for dunes. In the numerical model $\lambda_{\rm c}$ is close to $20\;l_0$ for a unidirectional wind regime \cite{Nart09}. Here, the angle between the primary wind and the dune alignment is $\alpha \sim 41^{\rm o}$, so that the minimum size $\lambda_{\rm c} \sin (\alpha)$ is approximately $13 \, l_0$. This value agrees with the observed width $W_0$. At the minimum size, the cross section has a dome dune shape with no slip face. Its transverse elevation profile looks like a parabola ($\phi=2/3$).


\section{Discussion and concluding remarks}

\subsection{From numerical outputs to field observations}

The steady-state shape of solitary linear dunes is characterized by self-similar cross sections with slip faces on both sides in the lower part, capped by a reworked part with a smaller aspect ratio. The width and height decrease linearly from the sand source to the dune tip where the dune size appears to be controlled by the minimum size for dunes. On Earth, the minimum observed dune size $\lambda_{\rm{c}}$ is about $10 \, {\rm{m}}$ as revealed by incipient dunes and superimposed bedforms \cite{Elbe05, Elbe08,Lanc96}. Using this length to set the value of $l_0$ in the model, we obtain $l_0=0.5\;{\rm m}$. 
For a volumetric flux at the source $J_{\rm{in}}= 4.1 \; l_0^{3}\, t_0^{-1}$ and a drift potential $q_{\rm sat}= 0.23 \; l_0^{2}\, t_0^{-1}$ \cite{Nart09}, we found that the dune at steady state has a maximum width of $55 \, l_0$ and a length of $1175 \, l_0$ (Figure \ref{fig: 1}c).
This dune would then correspond on Earth to a $590 \, {\rm{m}}$ long linear dune with a maximum width of $27 \; {\rm{m}}$. Such linear dunes of moderate size are often observed after a topographic obstacle where sand accumulates in the lee side (Figure \ref{fig: 1}a). Note that the obstacle is several times wider than the dune itself. With a typical drift potential of $30 \, \rm{m^2\, yr^{-1}}$ as measured in Niger \cite{Luca14}, the equivalent sediment influx would be of $275 \, {\rm{m^3 \, yr^{-1}}}$. A $100 \; {\rm{m}}$ wide obstacle could provide such an influx, catching a free sand flux of ${2.75\, \rm{m^2\, yr^{-1}}}$. This is much smaller than the drift potential, as expected in regions where dunes elongate on a non-erodible floor. We anticipate that remote measurements of length and width of isolated dunes at steady state could be used to estimate the free sand flux in zones of low sand availability on Earth and other planetary bodies.

In contrast to our numerical results in bidirectional wind regimes, linear dunes of all sizes in nature show defects and sometimes sinuous crest lines \cite{Rubi08, Bris00}. Moreover, their aspect ratio is substantially smaller, ${\sim}\,0.15$ \cite{Bris07,Livi89,Rubi09,Rosk11,Telf07}. Secondary flows have been proposed to explain the dune sinuosity \cite{Tsoa82}. For example, they could be produced by the deflection of the primary wind due to its acute angle with the orientation of the dune crest.  This mechanism is not present in our numerical model and therefore cannot be tested. However, the model can be used to investigate fluctuations in wind strength and orientation as well as variations in transport and sediment properties, other factors that could also explain the observed complexity of dune shapes \cite{Gao16,Gao18}.

It is worth noting that a linear dune with a sinuous crest has been simulated using a continuous numerical dune model when the dune is submitted to a bidirectional wind regime with a low frequency cycle \cite{Part09}. In such conditions, the dune migrates back and forth laterally. The dune sinuosity is then reminiscent of the transverse instability of dunes  \cite{courrech2015dune,guignier2013sand,Part11,reffet2010formation}. This instability does not occur in our numerical setup because the sand source maintains the upstream end of the linear dune, which does not migrate laterally except at the tip.

\subsection{Comparison between two elementary dune types: finger and barchan dunes}

An elongating linear dune, also described as a finger dune by \citeA{Cour14}, can be seen as an elementary dune type. It is a simple, non-compound dune that form on a non-erodible bed (Figure~\ref{fig: 1}). To this regard, it represents the counterpart in a multidirectional wind regime of the crescentic barchan dune. These two elementary dune types differ not only in their morphology but also in their dynamics and stability. The barchan dune propagates downstream, while the finger dune elongates and eventually remains static if no change of external conditions. When propagating, a solitary barchan continuously loses sand from the horns and possibly gains sand from an incoming free sand flux. The gain is proportional to the dune width but the loss is almost constant. The loss slightly increases with the barchan size but does not vanish for small barchans. As a result, the barchan equilibrium size, for which gain and loss balance, is unstable \cite{hersen2004corridors}. Apart from this unstable equilibrium, a barchan either keeps growing or disappears. It is the contrary for an isolated finger dune elongating from a source where the equilibrium size is asymptotically stable (i.e., attractive), as shown in Figure~\ref{fig: 1}e. Because the source limits the gain to one dune end, it is independent on the dune length while the loss does increase with the length. Unlike the barchan dune, sand loss is not restricted to the dune ends but also occurs all along the dune. This is due to crest reversals, a fundamental characteristic of finger dunes in multidirectional wind regimes.

\subsection{Interactions of elongating linear dunes}

In dryland environments, dunes are rarely isolated objects but are parts of large assemblies and interact. Those interactions are of two kinds: (i) long-range interactions through sand flux, loss and capture, and (ii) collisions. The long-range interaction tends to organize barchan dunes in chevrons, aligning a downstream dune with one horn of the dune upstream. The collisions between barchans split dunes, which regularizes their size and organizes the field into corridors \cite{Elbe05, Elbe08, genois2013agent,hersen2005collision, worman2013modeling}. Hence, massive barchan dunes are rare and seem restricted to the upstream frontier of the dune field where collisions are not effective \cite{genois2013spatial, genois2016out}. 


\begin{figure}[ht!]
\begin{center}
\includegraphics[width=1.0\linewidth]{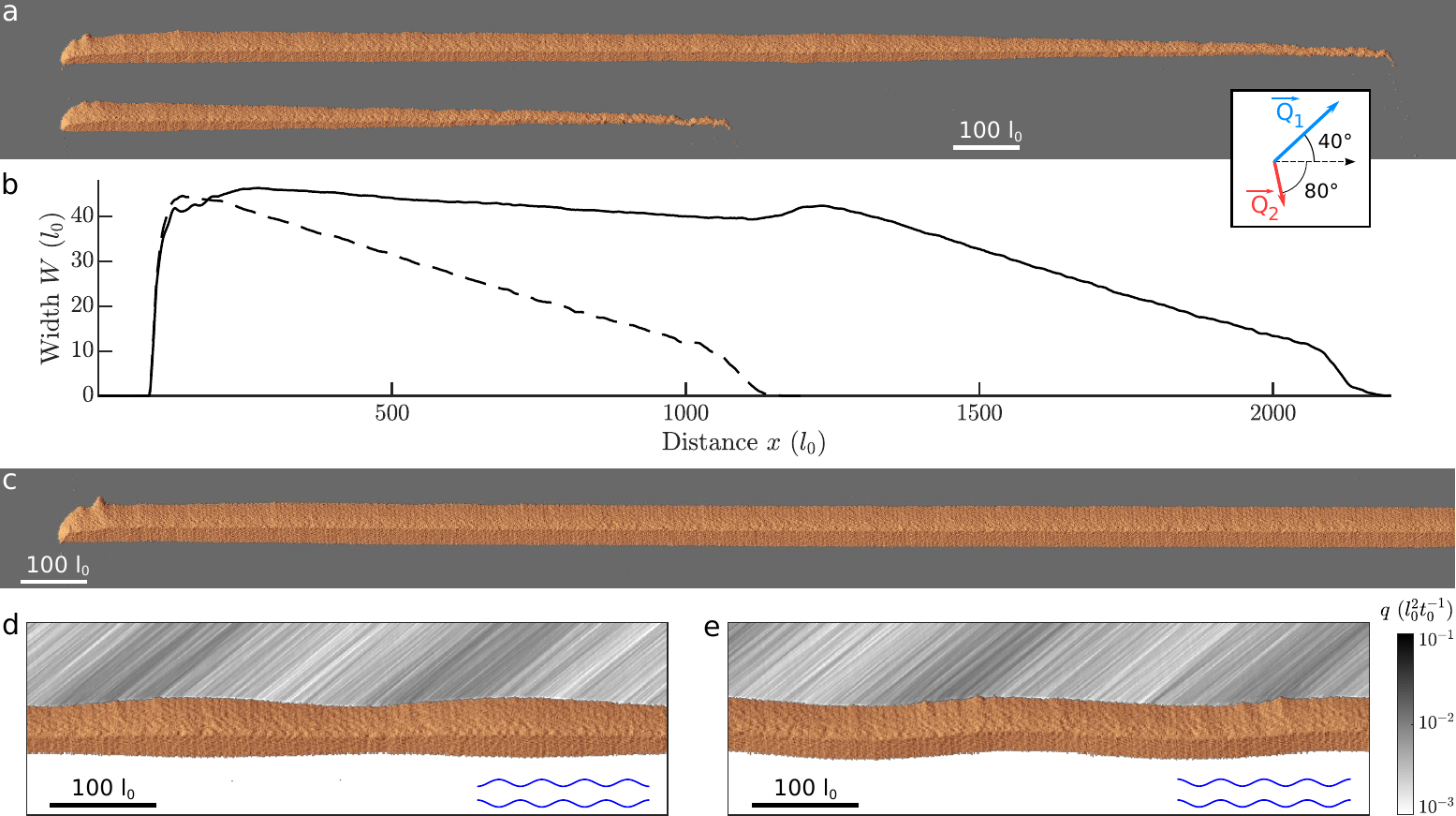}
\end{center}
\caption{
{\bf Interactions of elongating linear dunes.} 
\textbf{a:} Shapes at the end of the secondary wind of two parallel linear dunes at steady state. The two sources have an equal influx set to $3.2\,\, l_0^3\, t_0^{-1}$. 
\textbf{b:} Width of these two dunes with respect to distance (solid and dashed lines are for the long and short dunes, respectively).
\textbf{c:} Elongating dune with periodic boundary conditions perpendicularly to the direction of elongation and constant influx of $4.1\,\,l_0^3\, t_0^{-1}$. As sediment gains compensates for losses, the dune keeps elongating with a constant width.  
\textbf{d:} Outflux emitted by a linear dune with a straight crest line and a sinusoidal modulation of the cross-sectional area.
\textbf{e:} Outflux emitted by a linear dune with a sinuous body and a constant cross-sectional area. In \textbf{d} and \textbf{e}, the shape modulation impacts the outflux. Blue lines help to distinguish the respective shapes.
The inset on top shows the transport vectors of both winds used in simulations.
}
\label{fig: 4}
\end{figure}

Interactions between linear dunes remain to be studied, but we already outline some behaviors that illustrate the role of the longitudinal sediment budget. Figure~\ref{fig: 4}b shows a pair of steady linear dunes that have elongated from two separate sources with equal sediment influxes. Both dunes elongate side by side but, accordingly to the direction of the sand flux on a flat bed (Figure~\ref{fig: 1}c), the downstream dune catches the sand loss of the upstream one. The effective sediment influx of the downstream dune being twice the influx at the source, its steady length is twice the length of the upstream dune. As a result, the width and height of the downstream dune are almost constant when it is in the shadow of the upstream dune. The extra loss that occurs at the tip of the upstream dune locally increases the width of the downstream one. From this bulge, the width and height decrease with distance much like for the upstream dune, i.e., like for a solitary dune. In the absence of migration, these long-range interactions may regularize the width and height of elongating linear dunes within a field. However, without a net sediment loss along the dune, the length at steady state becomes virtually infinite and the stability of the equilibrium shape is compromised. This may explain that while giant barchan dunes are curiosities, fields of giant linear dunes are common in nature as observed on Earth or Titan.

Variations in dune shape can also affect the sand loss along linear dunes. Figure~\ref{fig: 4} shows the sand loss associated with a sinusoidal modulation of the cross-sectional area of a straight dune  (Figure~\ref{fig: 4}d), and with a sinuous body of constant cross-sectional area (Figure~\ref{fig: 4}e). In both cases, the sand loss remarkably reflects the topography and is inhomogeneous. It is maximum at the beginning of the dune enlargement (Figure \ref{fig: 4}d) or where the angle between the crest line and the primary wind is minimum (Figure~\ref{fig: 4}e), i.e., where the primary wind experiences a smaller aspect ratio. \citeA{Bris00} observed such an effect on a sinuous linear dune in the field.
Unlike for the idealized morphology of the elementary finger dune, sand loss is not constant anymore. The resulting heterogeneous mass exchange may pattern the entire dune field. 

When linear dunes elongate from fixed sources, mass exchange is limited to long-range interactions. But, when the source of sediment is not fixed or when the influx falls below a critical value, the linear dune can escape, migrate, and collide another dune. This migrating dune is already a hybrid of the two elementary dune types, the finger and the barchan dunes, and is sometimes referred to as an asymmetric barchan.

\section*{Acknowledgements}
We acknowledge support from the UnivEarthS LabEx program (ANR-10-LABX-0023), the IdEx Universit\'e de Paris (ANR-18-IDEX-0001), the French National Research Agency (ANR-17-CE01-0014/SONO) and the National Science Center of Poland (grant 2016/23/B/ST10/01700). Numerical simulations were partly performed on the S-CAPAD platform, IPGP, France. We used ReSCAL software (available at \url{http://www.ipgp.fr/rescal/rescal-linear-dune.tar.gz} with dedicated parameters) for all simulations. Wind data from ERA-Interim have been used to derive sand fluxes.


\begin{thebibliography}{}

\bibitem [\protect \citeauthoryear {%
Bristow%
, Augustinus%
, Wallis%
, Jol%
\BCBL {}\ \BBA {} Rhodes%
}{%
Bristow%
\ \protect \BOthers {.}}{%
{\protect \APACyear {2010}}%
}]{%
Bris10}
\APACinsertmetastar {%
Bris10}%
\begin{APACrefauthors}%
Bristow, C\BPBI S.%
, Augustinus, P\BPBI C.%
, Wallis, I.%
, Jol, H.%
\BCBL {}\ \BBA {} Rhodes, E.%
\end{APACrefauthors}%
\unskip\
\newblock
\APACrefYearMonthDay{2010}{}{}.
\newblock
{\BBOQ}\APACrefatitle {Investigation of the age and migration of reversing
  dunes in {A}ntarctica using {GPR} and {OSL}, with implications for {GPR} on
  {M}ars} {Investigation of the age and migration of reversing dunes in
  {A}ntarctica using {GPR} and {OSL}, with implications for {GPR} on
  {M}ars}.{\BBCQ}
\newblock
\APACjournalVolNumPages{Earth and Planetary Science Letters}{289}{1-2}{30--42}.
\PrintBackRefs{\CurrentBib}

\bibitem [\protect \citeauthoryear {%
{Bristow}%
, {Bailey}%
\BCBL {}\ \BBA {} {Lancaster}%
}{%
{Bristow}%
\ \protect \BOthers {.}}{%
{\protect \APACyear {2000}}%
}]{%
Bris00}
\APACinsertmetastar {%
Bris00}%
\begin{APACrefauthors}%
{Bristow}, C\BPBI S.%
, {Bailey}, S\BPBI D.%
\BCBL {}\ \BBA {} {Lancaster}, N.%
\end{APACrefauthors}%
\unskip\
\newblock
\APACrefYearMonthDay{2000}{{\APACmonth{07}}}{}.
\newblock
{\BBOQ}\APACrefatitle {{The sedimentary structure of linear sand dunes}} {{The
  sedimentary structure of linear sand dunes}}.{\BBCQ}
\newblock
\APACjournalVolNumPages{Nature}{406}{}{56-59}.
\PrintBackRefs{\CurrentBib}

\bibitem [\protect \citeauthoryear {%
Bristow%
, Duller%
\BCBL {}\ \BBA {} Lancaster%
}{%
Bristow%
\ \protect \BOthers {.}}{%
{\protect \APACyear {2007}}%
}]{%
Bris07}
\APACinsertmetastar {%
Bris07}%
\begin{APACrefauthors}%
Bristow, C\BPBI S.%
, Duller, G.%
\BCBL {}\ \BBA {} Lancaster, N.%
\end{APACrefauthors}%
\unskip\
\newblock
\APACrefYearMonthDay{2007}{}{}.
\newblock
{\BBOQ}\APACrefatitle {Age and dynamics of linear dunes in the Namib Desert}
  {Age and dynamics of linear dunes in the namib desert}.{\BBCQ}
\newblock
\APACjournalVolNumPages{Geology}{35}{6}{555--558}.
\PrintBackRefs{\CurrentBib}

\bibitem [\protect \citeauthoryear {%
Courrech~du Pont%
}{%
Courrech~du Pont%
}{%
{\protect \APACyear {2015}}%
}]{%
courrech2015dune}
\APACinsertmetastar {%
courrech2015dune}%
\begin{APACrefauthors}%
Courrech~du Pont, S.%
\end{APACrefauthors}%
\unskip\
\newblock
\APACrefYearMonthDay{2015}{}{}.
\newblock
{\BBOQ}\APACrefatitle {Dune morphodynamics} {Dune morphodynamics}.{\BBCQ}
\newblock
\APACjournalVolNumPages{Comptes Rendus Physique}{16}{1}{118--138}.
\PrintBackRefs{\CurrentBib}

\bibitem [\protect \citeauthoryear {%
Courrech~du Pont%
, Narteau%
\BCBL {}\ \BBA {} Gao%
}{%
Courrech~du Pont%
\ \protect \BOthers {.}}{%
{\protect \APACyear {2014}}%
}]{%
Cour14}
\APACinsertmetastar {%
Cour14}%
\begin{APACrefauthors}%
Courrech~du Pont, S.%
, Narteau, C.%
\BCBL {}\ \BBA {} Gao, X.%
\end{APACrefauthors}%
\unskip\
\newblock
\APACrefYearMonthDay{2014}{}{}.
\newblock
{\BBOQ}\APACrefatitle {Two modes for dune orientation} {Two modes for dune
  orientation}.{\BBCQ}
\newblock
\APACjournalVolNumPages{Geology}{42}{9}{743--746}.
\PrintBackRefs{\CurrentBib}

\bibitem [\protect \citeauthoryear {%
Elbelrhiti%
, Andreotti%
\BCBL {}\ \BBA {} Claudin%
}{%
Elbelrhiti%
\ \protect \BOthers {.}}{%
{\protect \APACyear {2008}}%
}]{%
Elbe08}
\APACinsertmetastar {%
Elbe08}%
\begin{APACrefauthors}%
Elbelrhiti, H.%
, Andreotti, B.%
\BCBL {}\ \BBA {} Claudin, P.%
\end{APACrefauthors}%
\unskip\
\newblock
\APACrefYearMonthDay{2008}{}{}.
\newblock
{\BBOQ}\APACrefatitle {Barchan dune corridors: field characterization and
  investigation of control parameters} {Barchan dune corridors: field
  characterization and investigation of control parameters}.{\BBCQ}
\newblock
\APACjournalVolNumPages{J. Geophys. Res. Earth Surf.}{113}{F2}{}.
\PrintBackRefs{\CurrentBib}

\bibitem [\protect \citeauthoryear {%
Elbelrhiti%
, Claudin%
\BCBL {}\ \BBA {} Andreotti%
}{%
Elbelrhiti%
\ \protect \BOthers {.}}{%
{\protect \APACyear {2005}}%
}]{%
Elbe05}
\APACinsertmetastar {%
Elbe05}%
\begin{APACrefauthors}%
Elbelrhiti, H.%
, Claudin, P.%
\BCBL {}\ \BBA {} Andreotti, B.%
\end{APACrefauthors}%
\unskip\
\newblock
\APACrefYearMonthDay{2005}{}{}.
\newblock
{\BBOQ}\APACrefatitle {Field evidence for surface-wave-induced instability of
  sand dunes} {Field evidence for surface-wave-induced instability of sand
  dunes}.{\BBCQ}
\newblock
\APACjournalVolNumPages{Nature}{437}{7059}{720}.
\PrintBackRefs{\CurrentBib}

\bibitem [\protect \citeauthoryear {%
Gadal%
, Narteau%
, du Pont%
, Rozier%
\BCBL {}\ \BBA {} Claudin%
}{%
Gadal%
\ \protect \BOthers {.}}{%
{\protect \APACyear {2019}}%
}]{%
Gada19}
\APACinsertmetastar {%
Gada19}%
\begin{APACrefauthors}%
Gadal, C.%
, Narteau, C.%
, du Pont, S\BPBI C.%
, Rozier, O.%
\BCBL {}\ \BBA {} Claudin, P.%
\end{APACrefauthors}%
\unskip\
\newblock
\APACrefYearMonthDay{2019}{}{}.
\newblock
{\BBOQ}\APACrefatitle {Incipient bedforms in a bidirectional wind regime}
  {Incipient bedforms in a bidirectional wind regime}.{\BBCQ}
\newblock
\APACjournalVolNumPages{Journal of Fluid Mechanics}{862}{}{490--516}.
\PrintBackRefs{\CurrentBib}

\bibitem [\protect \citeauthoryear {%
Gao%
, Gadal%
, Rozier%
\BCBL {}\ \BBA {} Narteau%
}{%
Gao%
\ \protect \BOthers {.}}{%
{\protect \APACyear {2018}}%
}]{%
Gao18}
\APACinsertmetastar {%
Gao18}%
\begin{APACrefauthors}%
Gao, X.%
, Gadal, C.%
, Rozier, O.%
\BCBL {}\ \BBA {} Narteau, C.%
\end{APACrefauthors}%
\unskip\
\newblock
\APACrefYearMonthDay{2018}{}{}.
\newblock
{\BBOQ}\APACrefatitle {Morphodynamics of barchan and dome dunes under variable
  wind regimes} {Morphodynamics of barchan and dome dunes under variable wind
  regimes}.{\BBCQ}
\newblock
\APACjournalVolNumPages{Geology}{46}{9}{743}.
\newblock
\begin{APACrefDOI} \doi{10.1130/G45101.1} \end{APACrefDOI}
\PrintBackRefs{\CurrentBib}

\bibitem [\protect \citeauthoryear {%
Gao%
, Narteau%
\BCBL {}\ \BBA {} Rozier%
}{%
Gao%
\ \protect \BOthers {.}}{%
{\protect \APACyear {2015}}%
}]{%
Gao15b}
\APACinsertmetastar {%
Gao15b}%
\begin{APACrefauthors}%
Gao, X.%
, Narteau, C.%
\BCBL {}\ \BBA {} Rozier, O.%
\end{APACrefauthors}%
\unskip\
\newblock
\APACrefYearMonthDay{2015}{}{}.
\newblock
{\BBOQ}\APACrefatitle {Development and steady states of transverse dunes: A
  numerical analysis of dune pattern coarsening and giant dunes} {Development
  and steady states of transverse dunes: A numerical analysis of dune pattern
  coarsening and giant dunes}.{\BBCQ}
\newblock
\APACjournalVolNumPages{Journal of Geophysical Research: Earth
  Surface}{120}{}{2200--2219}.
\PrintBackRefs{\CurrentBib}

\bibitem [\protect \citeauthoryear {%
Gao%
, Narteau%
\BCBL {}\ \BBA {} Rozier%
}{%
Gao%
\ \protect \BOthers {.}}{%
{\protect \APACyear {2016}}%
}]{%
Gao16}
\APACinsertmetastar {%
Gao16}%
\begin{APACrefauthors}%
Gao, X.%
, Narteau, C.%
\BCBL {}\ \BBA {} Rozier, O.%
\end{APACrefauthors}%
\unskip\
\newblock
\APACrefYearMonthDay{2016}{}{}.
\newblock
{\BBOQ}\APACrefatitle {Controls on and effects of armoring and vertical sorting
  in aeolian dune fields: a numerical simulation study} {Controls on and
  effects of armoring and vertical sorting in aeolian dune fields: a numerical
  simulation study}.{\BBCQ}
\newblock
\APACjournalVolNumPages{Geophys. Res. Lett.}{43}{}{2614--2622}.
\newblock
\begin{APACrefDOI} \doi{10.1002/2016GL068416} \end{APACrefDOI}
\PrintBackRefs{\CurrentBib}

\bibitem [\protect \citeauthoryear {%
Gao%
, Narteau%
, Rozier%
\BCBL {}\ \BBA {} Courrech~du Pont%
}{%
Gao%
\ \protect \BOthers {.}}{%
{\protect \APACyear {2015a}}%
}]{%
Gao15a}
\APACinsertmetastar {%
Gao15a}%
\begin{APACrefauthors}%
Gao, X.%
, Narteau, C.%
, Rozier, O.%
\BCBL {}\ \BBA {} Courrech~du Pont, S.%
\end{APACrefauthors}%
\unskip\
\newblock
\APACrefYearMonthDay{2015a}{}{}.
\newblock
{\BBOQ}\APACrefatitle {Phase diagrams of dune shape and orientation depending
  on sand availability} {Phase diagrams of dune shape and orientation depending
  on sand availability}.{\BBCQ}
\newblock
\APACjournalVolNumPages{Scientific reports}{5}{}{14677}.
\PrintBackRefs{\CurrentBib}

\bibitem [\protect \citeauthoryear {%
G{\'e}nois%
, Courrech~du Pont%
, Hersen%
\BCBL {}\ \BBA {} Gr{\'e}goire%
}{%
G{\'e}nois%
, Courrech~du Pont%
\BCBL {}\ \protect \BOthers {.}}{%
{\protect \APACyear {2013}}%
}]{%
genois2013agent}
\APACinsertmetastar {%
genois2013agent}%
\begin{APACrefauthors}%
G{\'e}nois, M.%
, Courrech~du Pont, S.%
, Hersen, P.%
\BCBL {}\ \BBA {} Gr{\'e}goire, G.%
\end{APACrefauthors}%
\unskip\
\newblock
\APACrefYearMonthDay{2013}{}{}.
\newblock
{\BBOQ}\APACrefatitle {An agent-based model of dune interactions produces the
  emergence of patterns in deserts} {An agent-based model of dune interactions
  produces the emergence of patterns in deserts}.{\BBCQ}
\newblock
\APACjournalVolNumPages{Geophysical Research Letters}{40}{15}{3909--3914}.
\PrintBackRefs{\CurrentBib}

\bibitem [\protect \citeauthoryear {%
G{\'e}nois%
, Hersen%
, Bertin%
, Courrech~du Pont%
\BCBL {}\ \BBA {} Gr{\'e}goire%
}{%
G{\'e}nois%
\ \protect \BOthers {.}}{%
{\protect \APACyear {2016}}%
}]{%
genois2016out}
\APACinsertmetastar {%
genois2016out}%
\begin{APACrefauthors}%
G{\'e}nois, M.%
, Hersen, P.%
, Bertin, E.%
, Courrech~du Pont, S.%
\BCBL {}\ \BBA {} Gr{\'e}goire, G.%
\end{APACrefauthors}%
\unskip\
\newblock
\APACrefYearMonthDay{2016}{}{}.
\newblock
{\BBOQ}\APACrefatitle {Out-of-equilibrium stationary states, percolation, and
  subcritical instabilities in a fully nonconservative system}
  {Out-of-equilibrium stationary states, percolation, and subcritical
  instabilities in a fully nonconservative system}.{\BBCQ}
\newblock
\APACjournalVolNumPages{Physical Review E}{94}{4}{042101}.
\PrintBackRefs{\CurrentBib}

\bibitem [\protect \citeauthoryear {%
G{\'e}nois%
, Hersen%
, Courrech~du Pont%
\BCBL {}\ \BBA {} Gr{\'e}goire%
}{%
G{\'e}nois%
, Hersen%
\BCBL {}\ \protect \BOthers {.}}{%
{\protect \APACyear {2013}}%
}]{%
genois2013spatial}
\APACinsertmetastar {%
genois2013spatial}%
\begin{APACrefauthors}%
G{\'e}nois, M.%
, Hersen, P.%
, Courrech~du Pont, S.%
\BCBL {}\ \BBA {} Gr{\'e}goire, G.%
\end{APACrefauthors}%
\unskip\
\newblock
\APACrefYearMonthDay{2013}{}{}.
\newblock
{\BBOQ}\APACrefatitle {Spatial structuring and size selection as collective
  behaviours in an agent-based model for barchan fields} {Spatial structuring
  and size selection as collective behaviours in an agent-based model for
  barchan fields}.{\BBCQ}
\newblock
\APACjournalVolNumPages{The European Physical Journal B}{86}{11}{447}.
\PrintBackRefs{\CurrentBib}

\bibitem [\protect \citeauthoryear {%
Guignier%
, Niiya%
, Nishimori%
, Lague%
\BCBL {}\ \BBA {} Valance%
}{%
Guignier%
\ \protect \BOthers {.}}{%
{\protect \APACyear {2013}}%
}]{%
guignier2013sand}
\APACinsertmetastar {%
guignier2013sand}%
\begin{APACrefauthors}%
Guignier, L.%
, Niiya, H.%
, Nishimori, H.%
, Lague, D.%
\BCBL {}\ \BBA {} Valance, A.%
\end{APACrefauthors}%
\unskip\
\newblock
\APACrefYearMonthDay{2013}{}{}.
\newblock
{\BBOQ}\APACrefatitle {Sand dunes as migrating strings} {Sand dunes as
  migrating strings}.{\BBCQ}
\newblock
\APACjournalVolNumPages{Physical Review E}{87}{5}{052206}.
\PrintBackRefs{\CurrentBib}

\bibitem [\protect \citeauthoryear {%
Hersen%
\ \protect \BOthers {.}}{%
Hersen%
\ \protect \BOthers {.}}{%
{\protect \APACyear {2004}}%
}]{%
hersen2004corridors}
\APACinsertmetastar {%
hersen2004corridors}%
\begin{APACrefauthors}%
Hersen, P.%
, Andersen, K\BPBI H.%
, Elbelrhiti, H.%
, Andreotti, B.%
, Claudin, P.%
\BCBL {}\ \BBA {} Douady, S.%
\end{APACrefauthors}%
\unskip\
\newblock
\APACrefYearMonthDay{2004}{}{}.
\newblock
{\BBOQ}\APACrefatitle {Corridors of barchan dunes: Stability and size
  selection} {Corridors of barchan dunes: Stability and size selection}.{\BBCQ}
\newblock
\APACjournalVolNumPages{Physical Review E}{69}{1}{011304}.
\PrintBackRefs{\CurrentBib}

\bibitem [\protect \citeauthoryear {%
Hersen%
\ \BBA {} Douady%
}{%
Hersen%
\ \BBA {} Douady%
}{%
{\protect \APACyear {2005}}%
}]{%
hersen2005collision}
\APACinsertmetastar {%
hersen2005collision}%
\begin{APACrefauthors}%
Hersen, P.%
\BCBT {}\ \BBA {} Douady, S.%
\end{APACrefauthors}%
\unskip\
\newblock
\APACrefYearMonthDay{2005}{}{}.
\newblock
{\BBOQ}\APACrefatitle {Collision of barchan dunes as a mechanism of size
  regulation} {Collision of barchan dunes as a mechanism of size
  regulation}.{\BBCQ}
\newblock
\APACjournalVolNumPages{Geophysical Research Letters}{32}{21}{}.
\PrintBackRefs{\CurrentBib}

\bibitem [\protect \citeauthoryear {%
Lancaster%
}{%
Lancaster%
}{%
{\protect \APACyear {1982}}%
}]{%
Lanc82}
\APACinsertmetastar {%
Lanc82}%
\begin{APACrefauthors}%
Lancaster, N.%
\end{APACrefauthors}%
\unskip\
\newblock
\APACrefYearMonthDay{1982}{}{}.
\newblock
{\BBOQ}\APACrefatitle {Linear dunes} {Linear dunes}.{\BBCQ}
\newblock
\APACjournalVolNumPages{Prog. Phys. Geogr.}{6}{4}{475--504}.
\PrintBackRefs{\CurrentBib}

\bibitem [\protect \citeauthoryear {%
Lancaster%
}{%
Lancaster%
}{%
{\protect \APACyear {1996}}%
}]{%
Lanc96}
\APACinsertmetastar {%
Lanc96}%
\begin{APACrefauthors}%
Lancaster, N.%
\end{APACrefauthors}%
\unskip\
\newblock
\APACrefYearMonthDay{1996}{}{}.
\newblock
{\BBOQ}\APACrefatitle {Field studies of sand patch initiation processes on the
  northern margin of the Namib sand sea} {Field studies of sand patch
  initiation processes on the northern margin of the namib sand sea}.{\BBCQ}
\newblock
\APACjournalVolNumPages{Earth Surface Processes and
  Landforms}{21}{10}{947--954}.
\PrintBackRefs{\CurrentBib}

\bibitem [\protect \citeauthoryear {%
Livingstone%
}{%
Livingstone%
}{%
{\protect \APACyear {1989}}%
}]{%
Livi89}
\APACinsertmetastar {%
Livi89}%
\begin{APACrefauthors}%
Livingstone, I.%
\end{APACrefauthors}%
\unskip\
\newblock
\APACrefYearMonthDay{1989}{}{}.
\newblock
{\BBOQ}\APACrefatitle {Monitoring surface change on a Namib linear dune}
  {Monitoring surface change on a namib linear dune}.{\BBCQ}
\newblock
\APACjournalVolNumPages{Earth Surface Processes and
  Landforms}{14}{4}{317--332}.
\PrintBackRefs{\CurrentBib}

\bibitem [\protect \citeauthoryear {%
L{\"u}%
, Narteau%
, Dong%
, Rozier%
\BCBL {}\ \BBA {} Courrech Du~Pont%
}{%
L{\"u}%
\ \protect \BOthers {.}}{%
{\protect \APACyear {2017}}%
}]{%
Lu17}
\APACinsertmetastar {%
Lu17}%
\begin{APACrefauthors}%
L{\"u}, P.%
, Narteau, C.%
, Dong, Z.%
, Rozier, O.%
\BCBL {}\ \BBA {} Courrech Du~Pont, S.%
\end{APACrefauthors}%
\unskip\
\newblock
\APACrefYearMonthDay{2017}{}{}.
\newblock
{\BBOQ}\APACrefatitle {Unravelling raked linear dunes to explain the
  coexistence of bedforms in complex dunefields} {Unravelling raked linear
  dunes to explain the coexistence of bedforms in complex dunefields}.{\BBCQ}
\newblock
\APACjournalVolNumPages{Nature Communications}{8}{}{14239}.
\PrintBackRefs{\CurrentBib}

\bibitem [\protect \citeauthoryear {%
Lucas%
\ \protect \BOthers {.}}{%
Lucas%
\ \protect \BOthers {.}}{%
{\protect \APACyear {2014}}%
}]{%
Luca14}
\APACinsertmetastar {%
Luca14}%
\begin{APACrefauthors}%
Lucas, A.%
, Rodriguez, S.%
, Narteau, C.%
, Charnay, B.%
, Pont, S\BPBI C.%
, Tokano, T.%
\BDBL {}others%
\end{APACrefauthors}%
\unskip\
\newblock
\APACrefYearMonthDay{2014}{}{}.
\newblock
{\BBOQ}\APACrefatitle {Growth mechanisms and dune orientation on {T}itan}
  {Growth mechanisms and dune orientation on {T}itan}.{\BBCQ}
\newblock
\APACjournalVolNumPages{Geophys. Res. Lett.}{41}{17}{6093--6100}.
\PrintBackRefs{\CurrentBib}

\bibitem [\protect \citeauthoryear {%
Mainguet%
\ \BBA {} Callot%
}{%
Mainguet%
\ \BBA {} Callot%
}{%
{\protect \APACyear {1978}}%
}]{%
Main78}
\APACinsertmetastar {%
Main78}%
\begin{APACrefauthors}%
Mainguet, M.%
\BCBT {}\ \BBA {} Callot, Y.%
\end{APACrefauthors}%
\unskip\
\newblock
\APACrefYear{1978}.
\newblock
\APACrefbtitle {L'Erg de Fachi-Bilma (Tchad-Niger): Contribution a la
  connaissance de la dynamique des ergs et des dunes des zones arides chaudes}
  {L'erg de fachi-bilma (tchad-niger): Contribution a la connaissance de la
  dynamique des ergs et des dunes des zones arides chaudes}.
\newblock
\APACaddressPublisher{}{{\'E}d. du Centre national de la recherche
  scientifique}.
\PrintBackRefs{\CurrentBib}

\bibitem [\protect \citeauthoryear {%
Narteau%
, Zhang%
, Rozier%
\BCBL {}\ \BBA {} Claudin%
}{%
Narteau%
\ \protect \BOthers {.}}{%
{\protect \APACyear {2009}}%
}]{%
Nart09}
\APACinsertmetastar {%
Nart09}%
\begin{APACrefauthors}%
Narteau, C.%
, Zhang, D.%
, Rozier, O.%
\BCBL {}\ \BBA {} Claudin, P.%
\end{APACrefauthors}%
\unskip\
\newblock
\APACrefYearMonthDay{2009}{}{}.
\newblock
{\BBOQ}\APACrefatitle {Setting the length and time scales of a cellular
  automaton dune model from the analysis of superimposed bed forms} {Setting
  the length and time scales of a cellular automaton dune model from the
  analysis of superimposed bed forms}.{\BBCQ}
\newblock
\APACjournalVolNumPages{J. Geophys. Res.}{114}{}{F03006}.
\newblock
\begin{APACrefDOI} \doi{10.1029/2008JF001127} \end{APACrefDOI}
\PrintBackRefs{\CurrentBib}

\bibitem [\protect \citeauthoryear {%
Parteli%
, Andrade~Jr%
\BCBL {}\ \BBA {} Herrmann%
}{%
Parteli%
\ \protect \BOthers {.}}{%
{\protect \APACyear {2011}}%
}]{%
Part11}
\APACinsertmetastar {%
Part11}%
\begin{APACrefauthors}%
Parteli, E\BPBI J.%
, Andrade~Jr, J\BPBI S.%
\BCBL {}\ \BBA {} Herrmann, H\BPBI J.%
\end{APACrefauthors}%
\unskip\
\newblock
\APACrefYearMonthDay{2011}{}{}.
\newblock
{\BBOQ}\APACrefatitle {Transverse instability of dunes} {Transverse instability
  of dunes}.{\BBCQ}
\newblock
\APACjournalVolNumPages{Physical Review Letters}{107}{18}{188001}.
\PrintBackRefs{\CurrentBib}

\bibitem [\protect \citeauthoryear {%
Parteli%
, Dur{\'a}n%
, Tsoar%
, Schw{\"a}mmle%
\BCBL {}\ \BBA {} Herrmann%
}{%
Parteli%
\ \protect \BOthers {.}}{%
{\protect \APACyear {2009}}%
}]{%
Part09}
\APACinsertmetastar {%
Part09}%
\begin{APACrefauthors}%
Parteli, E\BPBI J.%
, Dur{\'a}n, O.%
, Tsoar, H.%
, Schw{\"a}mmle, V.%
\BCBL {}\ \BBA {} Herrmann, H\BPBI J.%
\end{APACrefauthors}%
\unskip\
\newblock
\APACrefYearMonthDay{2009}{}{}.
\newblock
{\BBOQ}\APACrefatitle {Dune formation under bimodal winds} {Dune formation
  under bimodal winds}.{\BBCQ}
\newblock
\APACjournalVolNumPages{Proc. Natl. Acad. Sci. U. S.
  A.}{106}{52}{22085--22089}.
\PrintBackRefs{\CurrentBib}

\bibitem [\protect \citeauthoryear {%
Reffet%
, Courrech~du Pont%
, Hersen%
\BCBL {}\ \BBA {} Douady%
}{%
Reffet%
\ \protect \BOthers {.}}{%
{\protect \APACyear {2010}}%
}]{%
reffet2010formation}
\APACinsertmetastar {%
reffet2010formation}%
\begin{APACrefauthors}%
Reffet, E.%
, Courrech~du Pont, S.%
, Hersen, P.%
\BCBL {}\ \BBA {} Douady, S.%
\end{APACrefauthors}%
\unskip\
\newblock
\APACrefYearMonthDay{2010}{}{}.
\newblock
{\BBOQ}\APACrefatitle {Formation and stability of transverse and longitudinal
  sand dunes} {Formation and stability of transverse and longitudinal sand
  dunes}.{\BBCQ}
\newblock
\APACjournalVolNumPages{Geology}{38}{6}{491--494}.
\PrintBackRefs{\CurrentBib}

\bibitem [\protect \citeauthoryear {%
Roskin%
, Porat%
, Tsoar%
, Blumberg%
\BCBL {}\ \BBA {} Zander%
}{%
Roskin%
\ \protect \BOthers {.}}{%
{\protect \APACyear {2011}}%
}]{%
Rosk11}
\APACinsertmetastar {%
Rosk11}%
\begin{APACrefauthors}%
Roskin, J.%
, Porat, N.%
, Tsoar, H.%
, Blumberg, D\BPBI G.%
\BCBL {}\ \BBA {} Zander, A\BPBI M.%
\end{APACrefauthors}%
\unskip\
\newblock
\APACrefYearMonthDay{2011}{}{}.
\newblock
{\BBOQ}\APACrefatitle {Age, origin and climatic controls on vegetated linear
  dunes in the northwestern Negev Desert (Israel)} {Age, origin and climatic
  controls on vegetated linear dunes in the northwestern negev desert
  (israel)}.{\BBCQ}
\newblock
\APACjournalVolNumPages{Quaternary Science Reviews}{30}{13-14}{1649--1674}.
\PrintBackRefs{\CurrentBib}

\bibitem [\protect \citeauthoryear {%
Rozier%
\ \BBA {} Narteau%
}{%
Rozier%
\ \BBA {} Narteau%
}{%
{\protect \APACyear {2014}}%
}]{%
Rozi14}
\APACinsertmetastar {%
Rozi14}%
\begin{APACrefauthors}%
Rozier, O.%
\BCBT {}\ \BBA {} Narteau, C.%
\end{APACrefauthors}%
\unskip\
\newblock
\APACrefYearMonthDay{2014}{}{}.
\newblock
{\BBOQ}\APACrefatitle {A real-space cellular automaton laboratory} {A
  real-space cellular automaton laboratory}.{\BBCQ}
\newblock
\APACjournalVolNumPages{Earth Surf. Processes Landforms}{39}{1}{98--109}.
\PrintBackRefs{\CurrentBib}

\bibitem [\protect \citeauthoryear {%
Rubin%
\ \BBA {} Hesp%
}{%
Rubin%
\ \BBA {} Hesp%
}{%
{\protect \APACyear {2009}}%
}]{%
Rubi09}
\APACinsertmetastar {%
Rubi09}%
\begin{APACrefauthors}%
Rubin, D\BPBI M.%
\BCBT {}\ \BBA {} Hesp, P\BPBI A.%
\end{APACrefauthors}%
\unskip\
\newblock
\APACrefYearMonthDay{2009}{}{}.
\newblock
{\BBOQ}\APACrefatitle {Multiple origins of linear dunes on {E}arth and {T}itan}
  {Multiple origins of linear dunes on {E}arth and {T}itan}.{\BBCQ}
\newblock
\APACjournalVolNumPages{Nature Geoscience}{2}{}{653--658}.
\PrintBackRefs{\CurrentBib}

\bibitem [\protect \citeauthoryear {%
Rubin%
, Tsoar%
\BCBL {}\ \BBA {} Blumberg%
}{%
Rubin%
\ \protect \BOthers {.}}{%
{\protect \APACyear {2008}}%
}]{%
Rubi08}
\APACinsertmetastar {%
Rubi08}%
\begin{APACrefauthors}%
Rubin, D\BPBI M.%
, Tsoar, H.%
\BCBL {}\ \BBA {} Blumberg, D\BPBI G.%
\end{APACrefauthors}%
\unskip\
\newblock
\APACrefYearMonthDay{2008}{}{}.
\newblock
{\BBOQ}\APACrefatitle {A second look at western {S}inai seif dunes and their
  lateral migration} {A second look at western {S}inai seif dunes and their
  lateral migration}.{\BBCQ}
\newblock
\APACjournalVolNumPages{Geomorphology}{93}{3}{335--342}.
\PrintBackRefs{\CurrentBib}

\bibitem [\protect \citeauthoryear {%
Telfer%
\ \BBA {} Thomas%
}{%
Telfer%
\ \BBA {} Thomas%
}{%
{\protect \APACyear {2007}}%
}]{%
Telf07}
\APACinsertmetastar {%
Telf07}%
\begin{APACrefauthors}%
Telfer, M.%
\BCBT {}\ \BBA {} Thomas, D.%
\end{APACrefauthors}%
\unskip\
\newblock
\APACrefYearMonthDay{2007}{}{}.
\newblock
{\BBOQ}\APACrefatitle {Late Quaternary linear dune accumulation and
  chronostratigraphy of the southwestern Kalahari: implications for aeolian
  palaeoclimatic reconstructions and predictions of future dynamics} {Late
  quaternary linear dune accumulation and chronostratigraphy of the
  southwestern kalahari: implications for aeolian palaeoclimatic
  reconstructions and predictions of future dynamics}.{\BBCQ}
\newblock
\APACjournalVolNumPages{Quaternary Science Reviews}{26}{19-21}{2617--2630}.
\PrintBackRefs{\CurrentBib}

\bibitem [\protect \citeauthoryear {%
Tsoar%
}{%
Tsoar%
}{%
{\protect \APACyear {1982}}%
}]{%
Tsoa82}
\APACinsertmetastar {%
Tsoa82}%
\begin{APACrefauthors}%
Tsoar, H.%
\end{APACrefauthors}%
\unskip\
\newblock
\APACrefYearMonthDay{1982}{}{}.
\newblock
{\BBOQ}\APACrefatitle {Internal structure and surface geometry of longitudinal
  (seif) dunes} {Internal structure and surface geometry of longitudinal (seif)
  dunes}.{\BBCQ}
\newblock
\APACjournalVolNumPages{Journal of Sedimentary Research}{52}{3}{}.
\PrintBackRefs{\CurrentBib}

\bibitem [\protect \citeauthoryear {%
Tsoar%
}{%
Tsoar%
}{%
{\protect \APACyear {1989}}%
}]{%
Tsoa89}
\APACinsertmetastar {%
Tsoa89}%
\begin{APACrefauthors}%
Tsoar, H.%
\end{APACrefauthors}%
\unskip\
\newblock
\APACrefYearMonthDay{1989}{}{}.
\newblock
{\BBOQ}\APACrefatitle {Linear dunes-forms and formation} {Linear dunes-forms
  and formation}.{\BBCQ}
\newblock
\APACjournalVolNumPages{Progress in Physical Geography}{13}{4}{507--528}.
\PrintBackRefs{\CurrentBib}

\bibitem [\protect \citeauthoryear {%
Worman%
, Murray%
, Littlewood%
, Andreotti%
\BCBL {}\ \BBA {} Claudin%
}{%
Worman%
\ \protect \BOthers {.}}{%
{\protect \APACyear {2013}}%
}]{%
worman2013modeling}
\APACinsertmetastar {%
worman2013modeling}%
\begin{APACrefauthors}%
Worman, S\BPBI L.%
, Murray, A\BPBI B.%
, Littlewood, R.%
, Andreotti, B.%
\BCBL {}\ \BBA {} Claudin, P.%
\end{APACrefauthors}%
\unskip\
\newblock
\APACrefYearMonthDay{2013}{}{}.
\newblock
{\BBOQ}\APACrefatitle {Modeling emergent large-scale structures of barchan dune
  fields} {Modeling emergent large-scale structures of barchan dune
  fields}.{\BBCQ}
\newblock
\APACjournalVolNumPages{Geology}{41}{10}{1059--1062}.
\PrintBackRefs{\CurrentBib}

\bibitem [\protect \citeauthoryear {%
Zhang%
, Narteau%
\BCBL {}\ \BBA {} Rozier%
}{%
Zhang%
\ \protect \BOthers {.}}{%
{\protect \APACyear {2010}}%
}]{%
Zhan10}
\APACinsertmetastar {%
Zhan10}%
\begin{APACrefauthors}%
Zhang, D.%
, Narteau, C.%
\BCBL {}\ \BBA {} Rozier, O.%
\end{APACrefauthors}%
\unskip\
\newblock
\APACrefYearMonthDay{2010}{}{}.
\newblock
{\BBOQ}\APACrefatitle {Morphodynamics of barchan and transverse dunes using a
  cellular automaton model} {Morphodynamics of barchan and transverse dunes
  using a cellular automaton model}.{\BBCQ}
\newblock
\APACjournalVolNumPages{J. Geophys. Res.}{115}{F3}{F03041}.
\PrintBackRefs{\CurrentBib}

\bibitem [\protect \citeauthoryear {%
Zhang%
, Narteau%
, Rozier%
\BCBL {}\ \BBA {} Courrech~du Pont%
}{%
Zhang%
\ \protect \BOthers {.}}{%
{\protect \APACyear {2012}}%
}]{%
Zhan12}
\APACinsertmetastar {%
Zhan12}%
\begin{APACrefauthors}%
Zhang, D.%
, Narteau, C.%
, Rozier, O.%
\BCBL {}\ \BBA {} Courrech~du Pont, S.%
\end{APACrefauthors}%
\unskip\
\newblock
\APACrefYearMonthDay{2012}{}{}.
\newblock
{\BBOQ}\APACrefatitle {Morphology and dynamics of star dunes from numerical
  modelling} {Morphology and dynamics of star dunes from numerical
  modelling}.{\BBCQ}
\newblock
\APACjournalVolNumPages{Nature Geoscience}{5}{7}{463--467}.
\PrintBackRefs{\CurrentBib}

\bibitem [\protect \citeauthoryear {%
Zhang%
, Yang%
, Rozier%
\BCBL {}\ \BBA {} Narteau%
}{%
Zhang%
\ \protect \BOthers {.}}{%
{\protect \APACyear {2014}}%
}]{%
Zhan14}
\APACinsertmetastar {%
Zhan14}%
\begin{APACrefauthors}%
Zhang, D.%
, Yang, X.%
, Rozier, O.%
\BCBL {}\ \BBA {} Narteau, C.%
\end{APACrefauthors}%
\unskip\
\newblock
\APACrefYearMonthDay{2014}{}{}.
\newblock
{\BBOQ}\APACrefatitle {Mean sediment residence time in barchan dunes} {Mean
  sediment residence time in barchan dunes}.{\BBCQ}
\newblock
\APACjournalVolNumPages{J. Geophys. Res.}{119}{}{451--463}.
\newblock
\begin{APACrefDOI} \doi{10.1002/2013JF002833} \end{APACrefDOI}
\PrintBackRefs{\CurrentBib}

\end{thebibliography}

\end{document}